\documentclass[10pt]{iopart}
\usepackage{graphicx}
\def\no{\noindent}
\def\L{\Lambda}
\def\a{\alpha}
\def\s{\sigma}

\def\bchi{{\bar\chi}}
\def\bvphi{{\bar\varphi}}

\def\vphi{\varphi}
\def\beq{\begin{equation}}
\def\eeq{\end{equation}}

\begin{document}

\title{Correlations in Systems of Complex Directed Macromolecules}
\author{Ch. Moseley and K. Ziegler}
\address{Institut f\"ur Physik, Universit\"at Augsburg, Germany}

\centerline{Special Issue of Journal of Physics, Condensed Matter}
\centerline{``{\bf Polymers and Complex Matter}'' (tentative title)}
\centerline{- dedicated to the 60th Birthday of Prof. Lothar Sch\"afer}

\begin{abstract}
An ensemble of directed macromolecules on a lattice is considered, where the
constituting molecules are chosen randomly with $N$ different
colors. Molecules of the same color experience a hard-core (exclusion)
interaction. We study the robustness of the macromolecules with respect
to breaking and changing the color of a constituting molecule, using
a $1/N$ expansion. The properties depend strongly on the density of
macromolecules. In particular, the macromolecules are robust against
breaking and changing the color at high densities but fragile at intermediate
densities.
\end{abstract}

\maketitle

\section{Introduction}

In a liquid of molecules long chain-like macromolecules (MM) can be formed
due to a special steric intermolecular interaction. These MM are often 
called polymers \cite{degennes}. We assume that the MM are directed,
i.e. they have a preferred direction in the $d$-dimensional space. This
implies that they are stretched and do not form loops or overhangs.

In many systems the constituting molecules (CM) of the MM are 
identical. However, in more complex systems, like in biological substances, 
the CM are chemically different. An example for the latter are the 
sequences of amino acids in proteins \cite{li04,banavar03,banavar04} and
molecular sequences along a DNA helix. For our statistical model we assume 
that there are $N$ different kinds of molecules, represented by $N$
colors.
\begin{figure} 
\begin{center}
\includegraphics[scale=0.4]{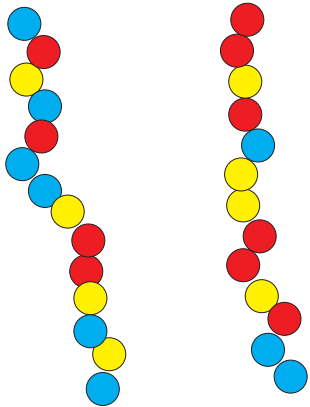}
\end{center} 
\caption{Two directed macromolecules.}
\end{figure}
They are randomly chosen to contribute to the MM such that the
resulting MM can be viewed as a chain of colored beads (cf. Fig. 1). 
An ensemble of MM can be considered as being created in a liquid of 
molecules by a steric interaction among the CM. As a result, we have a 
space-filling system of molecules, where some are constituting the MM and 
others exist as individual molecules. Properties of the ensemble on large
scales should not depend on local structures. This allows us to consider
a lattice model \cite{banavar03}, 
where each lattice site is occupied by a molecule of each
type. It is assumed that molecules of the same type have
a strong repulsive interaction. This restricts us to
at most one molecule of each type at any given lattice site.  
The fact that the interaction of different molecule types is neglected
can be understood as a chemical property of the molecules.

Any pair of neighboring molecules can form a bond state, representing
a pair of molecules, which
acts as a building block for an MM. There is a tendency of these bond
states to bond with other molecules. This leads eventually to
an equilibrium state, where we have a mixture of only large MM and 
individual molecules. The repulsive interaction between the molecules
creates correlations among the MM which lead to interesting properties
of the ensemble. For instance, the density of MM as a function of the
probability of a local bond state is of interest, as well as the question
how likely it is to break an MM or to replace a CM of special type
by one of a different type. The latter can happen during a collision process
when an individual molecule collides with an MM and exchanges with
a CM of the MM. 

\no
We consider an equilibrium ensemble of lattice MM and define three 
quantities to describe its global properties
with respect to the molecular structure.
There are two quantities that measure the correlations in real space. 
One measures how likely it is to break a directed MM 
and take it apart in space (cf. Fig. 2), the other one measures how 
likely it is to create a new (finite) MM by changing the type of the 
CM in an MM. If the ends of the broken MM or the ends of the new (finite) 
MM are $x$ and $x'$ we write for the corresponding quantities $G_{xx'}$ 
and $H_{xx'}$, respectively. It will be shown that $G_{xx'}$ and $H_{xx'}$ 
are directly related.
\begin{figure} 
\begin{center}
\includegraphics[scale=0.4]{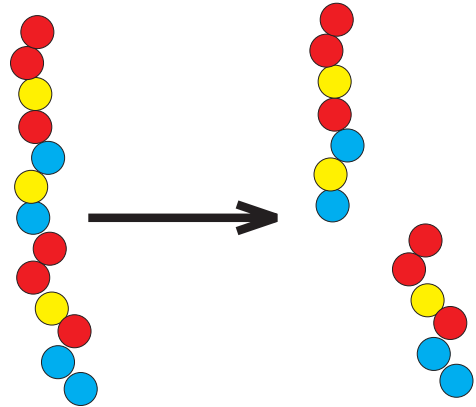}
\end{center} 
\caption{Breaking of a macromolecule.}
\end{figure}
Another quantity is the coherence in $\a$ (molecular-component) space.
This is called $S_{\a\a'}$ and  
measures how likely it is to replace a CM of type $\a$ in an MM
by one of type $\a'$ (cf. Fig. 3). 
\begin{figure} 
\begin{center}
\includegraphics[scale=0.4]{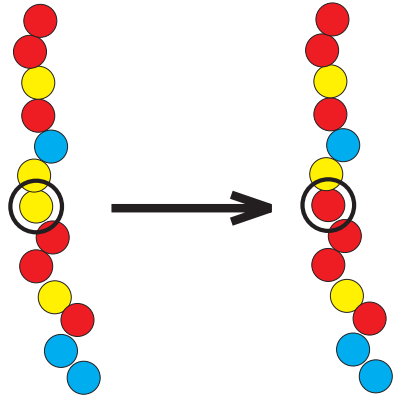}
\end{center} 
\caption{Changing the type of a constituting molecule inside a macromolecule,
caused by a collision with a molecule from the surrounding liquid.}
\end{figure}
All these quantities are evaluated in the limit $N\to\infty$ and
in a $1/N$ expansion.
We apply a functional-integral formalism developed in Ref. \cite{ziegler} 
to perform the calculations in a straightforward manner. 

\section{The Model}

A statistical model for the equilibrium statistics of directed MM was 
introduced in Ref. \cite{ziegler}. We briefly summarize the construction 
of the model and then turn to the evaluation of the interesting correlations
mentioned in the Introduction. The model is based on the idea that on each
site of a (cubic) lattice $N$ different molecules are located. They exist
either as individual (separated) molecules of a liquid with relative weight
$\mu$ or contribute to a directed MM as a bond state of a 
nearest-neighbor molecular pair at $x,x'$ of type $\a$ and $\a'$.
Moreover, the MM are directed, i.e. there is a special direction $t$ in the
$d$-dimensional space as a preferred direction of the MM. The fluctuations
of the MM are perpendicular to this, indicated by the $(d-1)$-dimensional
vector $r$. Therefore, we write $x=(t,r)$ for a site on the lattice.
The relative weight of a nearest-neighbor molecular pair 
at $x,x'$ of type $\a$ and $\a'$ is
$$
w_{x,x'}^{\a\a'}={1\over N}w_{x,x'}=\cases{
J/N & if $t'=t+1$, $r=r'$ or $r,r'$ nearest neighbors \cr
0 & otherwise 
}.
$$
The directed chain-like structure is given by the matrix $w$ that
is asymmetric in the direction $t$ but symmetric with respect to $r$.
$x'$ is assumed to be 'later' along the direction of the MM. 
$w$ is also symmetric with repect 
to $\a,\a'$. Thus $\a$ and $r$ are of the same type in contrast to $t$.

The statistical distribution of configurations of MM has to take into
account a strong repulsive interaction between molecules
of the same type at the same site. This can be conveniently be described
by a commutative algebra of nilpotent variables $\eta_x^{\a,\s}$ 
(i.e., $(\eta_x^{\a,\s})^l=0$ if $l>1$) with
$\a=1,2,...,N$ and $\s=1,2$ \cite{ziegler}.
Each molecule, characterized by its position $x=(r,t)$ on the lattice and 
by its chemical character $(\a,\s)$, is described by
a variable. Since $\{\eta_x^{\a,\s}\}$ are
nilpotent, the repulsive interaction
results in an exclusion principle for molecules of the same type. The atoms 
are considered as classical 
objects. Consequently, products of $\eta_x^{\a,\s}$ must be commutative. 
A segment of the MM (i.e. a bond state of two molecules or a 
dimer) reads
$$
w_{x,x'}^{\a\a'}\eta_x^{\a,1}\eta_{x'}^{\a',2}.
$$
A linear mapping of the algebra $\{\eta_x^{\a,\s}\}$ to the complex numbers
can be written as an integral on a lattice $\Lambda$ with \cite{berezin} 
$$
\int\prod_{x\in\L '\subseteq\L}\prod_{\a\in i_x}\prod_{\s\in j_{x,\a}}
\eta_x^{\a,\s}=\cases{
1 & if $\L '= \L$, $i_x=1,2,...,N$, $j_{x,\a}=1,2$\cr
0& otherwise\cr
}.
$$
The integral vanishes if the product is incomplete with respect to the
lattice $\Lambda$, to $\a=1,2,...,N$ or to $\sigma=1,2$. It can be used 
to write the statistical weight $P_I$ of a configuration $I$ of MM as
$$
P_I={1\over Z}\int W_I
$$
with
$$
W_I=\prod_{(x,\a;x',\a')\in I}{1\over N}w_{x,x'}\eta_x^{\a,1}
\eta_{x'}^{\a',2}\prod_{(x,\a)}(1+\mu\eta_x^{\a,1}\eta_x^{\a,2}),
$$
where $(x,\a;x',\a')$ is a bond between two CM of type $\a$ and $\a'$
at sites $x$ and $x'$.
The normalization $Z$ is the sum over all possible configurations
$\{ I\}$:
$$
Z=\sum_{\{ I\} }\int W_I .
$$
Finally, we impose periodic boundary conditions on the lattice. 
This implies a system on a torus with closed loops of MM. However,
the choice of boundary conditions should not be crucial for the
local properties of the MM like the density of the MM, $G_{xx'}$,
$H_{xx'}$, or $S_{\a\a'}$. 

\section{Functional-Integral Representation}

Introducing analytic functions of the nilpotent field and using the
properties of the integral, we may write for the normalization
\beq
Z=\sum_{\{ I\}}\int W_I =\int
\exp\Big[\sum_{x,x'}\sum_{\a,\a'=1}^N({1\over N}
w_{x,x'}+\mu\delta_{x,x'}\delta_{\a,\a'})\eta_x^{\a,1}\eta_{x'}^{\a',2}
\Big] .
\label{norm1}
\eeq
In other words, we can replace under the integral
$$
\sum_{\{I\}}W_I\to W=
\exp\Big[\sum_{x,x'}\sum_{\a,\a'=1}^N({1\over N}
w_{x,x'}+\mu\delta_{x,x'}\delta_{\a,\a'})\eta_x^{\a,1}\eta_{x'}^{\a',2}
\Big]
$$
in $Z$.
Another important consequence of the properties of the nilpotent field is
that we can calculate the probability that a point $x,\a$ belongs to a MM 
configuration $I$:
$$
{1\over Z}\int (1-\mu\eta_x^{\a,1}\eta_x^{\a,2})W_I=\cases{
P_I &if $(x,\a)\in I$\cr
0& otherwise\cr
}.
$$
After summation over all configurations $\{ I\}$, the probability
of a molecule of type $\a$ at site $x$ being part of any MM is
$$
\sum_{\{I\}}P_I=
%\langle (1-\mu\eta_x^{\a,1}\eta_x^{\a,2}) \rangle\equiv
\sum_{\{ I\} }{1\over Z}\int (1-\mu\eta_x^{\a,1}\eta_x^{\a,2})W_I
={1\over Z}\int (1-\mu\eta_x^{\a,1}\eta_x^{\a,2})W.
$$
This is also the local density $n(\mu)$ of an MM. In a similar manner,
we can express the correlations in terms of normalized integrals 
$$
\langle ... \rangle={1\over Z}\int...W
%\exp\Big[  \sum_{x,x'}\sum_{\a,\a'=1}^N({1\over N}
%w_{x,x'}+\mu\delta_{x,x'}\delta_{\a,\a'})\eta_x^{\a,1}\eta_{x'}^{\a',2}
%\Big]
$$
as correlations in real space  
$$
G_{xx'}
=\langle \eta_x^{\a,1}\eta_{x'}^{\a,2}\rangle
%=-\langle\psi_x^{\a,1}\bpsi_x^{\a,1}\bpsi_{x'}^{\a,2}\psi_{x'}^{\a,2}\rangle
=-C(x\a;x'\a),
$$
$$
H_{xx'}=\langle \eta_x^{\a,2}\eta_{x'}^{\a,1}\rangle
%=-\langle\psi_x^{\a,2}\bpsi_x^{\a,2}\bpsi_{x'}^{\a,1}\psi_{x'}^{\a,1}\rangle
=-C(x'\a;x\a),
$$
and correlations in $\a$ space   
$$
S_{\a\a'}=\langle\eta_x^{\a,1}\eta_{x}^{\a',2}\rangle
%=-\langle\psi_x^{\a,1}\bpsi_x^{\a,1}\bpsi_{x}^{\a',2}\psi_{x}^{\a',2}\rangle
=-C(x\a;x\a').
$$
These are correlation functions in real space and in $\a$ space,
respectively. The density of MM can also be expressed as
$$
n(\mu)=1 - \mu C(x\alpha;x\alpha).
$$
 
A lengthy but straightforward calculation \cite{ziegler} shows that
the normalization $Z$ in Eq. (\ref{norm1}) can be written as an
functional integral with respect to complex fields $\vphi$ and
$\chi$ as
\beq
Z=\int e^{-NS}\prod_x d\vphi_x d\chi_x.
\label{part2}
\eeq
with
$$
S=(\vphi,(1+w)^{-1}\bvphi)+(\chi,\bchi)-\sum_x
\log [\mu +(\vphi_x+i\chi_x)(\bvphi_x+i\bchi_x)].
$$
An analogous calculation for the correlation functions leads to
$$
\langle ...\rangle
={1\over Z}\int ...\ \  e^{-NS}\prod_x d\vphi_x d\chi_x.
\label{expect}
$$
The correlation functions then become
\begin{eqnarray*}
 && C(x\alpha ; x\alpha) =
 \left\langle \left[\mu+(\vphi_x+i\chi_x)(\vphi_x^\ast+i\chi_x^\ast)
\right]^{-1} \right\rangle \\
 && C(x\alpha ; x'\beta) =\mu
 \left\langle \frac{(\vphi_x+i\chi_x)(\vphi_{x'}^\ast+i\chi_{x'}^\ast)}{
 [\mu+(\vphi_x+i\chi_x)(\vphi_x^\ast+i\chi_x^\ast)]
 [\mu+(\vphi_{x'}+i\chi_{x'})(\vphi_{x'}^\ast+i\chi_{x'}^\ast)]} \right\rangle
% , (x,\alpha)\neq (x',\beta)
\end{eqnarray*}
for $(x,\alpha)\neq (x',\beta)$.

\subsection{$1/N$ Expansion}

The integrals in Eqs. (\ref{part2}) and (\ref{expect}) depend on the 
number of different molecule types
$N$ only through the prefactor $N$ in the exponential. This enables us to 
perform a saddle-point integration for large values of $N$ \cite{ziegler}. The
result is a $1/N$ expansion. The homogoneous saddle point of the integral in
Eq. (\ref{part2}) satisfies the relation
$$
i\chi =-{1\over 2}\vphi ,\ \ \ \ i\bchi =-{1\over 2}\bvphi.
$$
Then there is a trivial solution $\vphi_0 =\bvphi_0=\chi_0=\bchi_0=0$ and a
non--trivial solution with
$$
\vphi_1\bvphi_1=4(1-\mu).
$$
An expansion of the exponent in $Z$ in terms of the deviations around
the saddle point
$$
\phi_x=\vphi+i\chi+\delta\phi_x\ \ \ {\rm with}\ \
\delta\phi_x:=\delta\varphi_x + i\delta\chi_x,\ \ \ 
\delta\phi_x^\ast :=\delta\varphi_x^\ast + i\delta\chi_x^\ast
$$
provides corrections in powers of $1/N$.
For a non-zero density of MM (i.e. $\zeta\equiv\mu^{-1}>1$), an 
expansion up to second order leads to
$$
 C(x\alpha;x'\beta) \approx \zeta^{-1}-\zeta^{-2}+
 \frac 1{\zeta^3} [ -(\zeta-1)\langle \delta\phi_x^2 \rangle -
 2(\zeta-1)\langle \delta\phi_x \delta\phi_x^\ast \rangle -
 (\zeta-1)\langle \delta\phi_x\delta\phi_{x'} \rangle+
 \langle \delta\phi_x\delta\phi_{x'}^\ast \rangle
$$
$$
 +\sqrt{\zeta(\zeta-1)} \langle \delta\phi_x \rangle +
 (\zeta-1)^2 \langle (\delta\phi_x^\ast)^2 \rangle +
 (\zeta-1)^2 \langle \delta\phi_x^\ast \delta\phi_{x'} \rangle -
 (\zeta-1) \langle \delta\phi_x^\ast \delta\phi_{x'}^\ast \rangle
$$
$$
-(\zeta-1) \sqrt{\zeta(\zeta-1)} \langle \delta\phi_x \rangle +
 (\zeta-1)^2 \langle \delta\phi_{x'}^2 \rangle -
 2(\zeta-1) \langle \delta\phi_{x'} \delta\phi_{x'}^\ast \rangle -
 (\zeta-1) \sqrt{\zeta(\zeta-1)} \langle \delta\phi_{x'} \rangle
$$
$$
- (\zeta-1) \langle (\delta\phi_{x'}^\ast)^2 \rangle +
 \sqrt{\zeta(\zeta-1)} \langle \delta\phi_{x'}^\ast \rangle ],
$$
The first term is the large-$N$ limit, all other terms either vanish because of
$\langle\delta\phi_{x}\rangle =0$ or are of order $1/N$ because of
$\langle\delta\phi_{x'}^2 \rangle=o(1/N)$.
The saddle-point calculation leads to the result that the first order terms
vanish, and the second order terms can be expressed by the integrals
$$
%\begin{displaymath}
 \Phi(r-r',t-t'):= \left\langle \delta\phi_x \delta\phi_{x'} \right\rangle =
 \left\langle \delta\phi_x^\ast \delta\phi_{x'}^\ast \right\rangle \approx
% \hspace{0.5cm}
%\end{displaymath}
$$
$$
%\begin{displaymath}
% \hspace{0.5cm}
 \frac 1N \int  \int_0^{2\pi}
 \frac{B(k)^2 (\zeta-1)}{2B(k)\cos(\omega)-\zeta+(\zeta-2)B(k)^2}
 \, e^{i[k(r-r')-\omega(t-t')]}\frac{d\omega}{2\pi}\frac{d^{d-1}k}
{(2\pi)^{d-1}}
%\end{displaymath}
\label{corr1}
$$
$$
%\begin{displaymath}
 \hat\Phi(r-r',t-t'):= \left\langle
\delta\phi_x \delta\phi_{x'}^\ast \right\rangle \approx
% \hspace{0.5cm}
%\end{displaymath}
$$
$$
%\begin{displaymath}
% \hspace{0.5cm}
  \frac 1N \int  \int_0^{2\pi}
  \frac{B(k)(B(k)-\zeta\cos(\omega))}{2B(k)\cos(\omega)
   -\zeta+(\zeta-2)B(k)^2} e^{i[k(r-r')-\omega(t-t')]}
    \frac{d\omega}{2\pi}\frac{d^{d-1}k}{(2\pi)^{d-1}}
%\end{displaymath}
\label{corr2}
$$
with the function $B(k):=1-J+\frac J{d-1} \sum_{j=1}^{d-1} \cos k_j$.

\begin{figure}
\begin{center}
\includegraphics[scale=0.8]{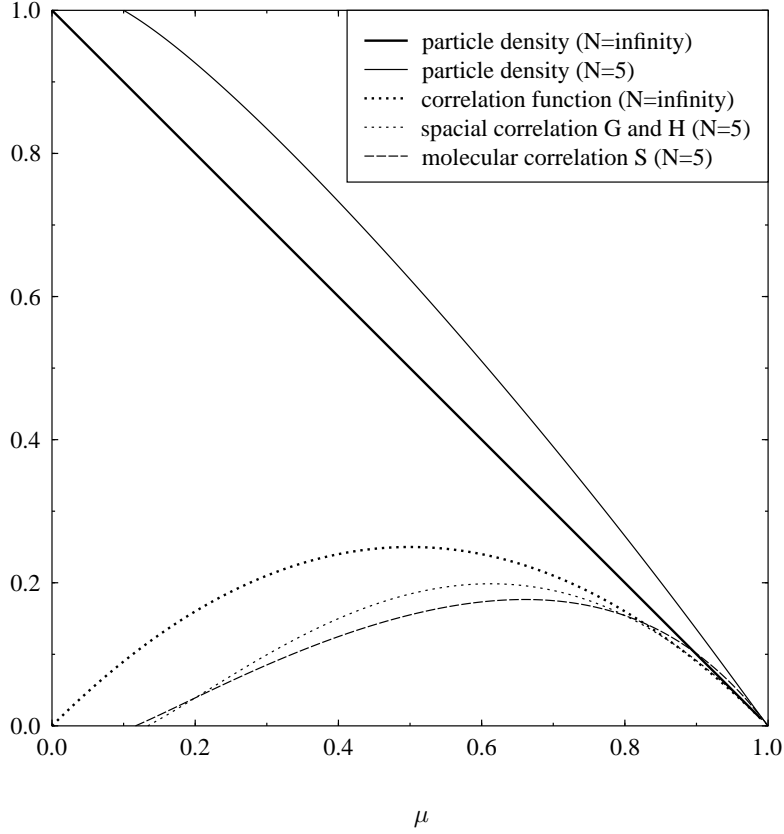}
\end{center}
\caption{The density of macromolecules and the correlations for
breaking a macromolecule ($G$) and replacing a molecule in a
macromolecule ($S$).}
\end{figure}

\section{Discussion of the Results}

The density of MM is
$$
n(\mu)=\cases{
1-\mu+o(1/N) & for $\mu<1$ \cr
0 & for $\mu>1$\cr
}.
$$
It increases with decreasing $N$, as shown in Fig. 4. This means that
it is more likely to form a MM if the number of available types of molecules
is reduced. 

How likely it is to break a MM depends on the position of the remaining ends
$x-x'$. The corresponding quantity $G_{xx'}$ decays with $|x-x'|$ which 
indicates that the molecules have a tendency to form a MM. However, it does
not vanish asymptotically but stays nonzero for $\zeta>1$ and $t'=t$: 
$$
G= \lim_{|r-r'|\rightarrow\infty} C(x\alpha;x'\alpha) \approx
(1-\zeta^{-1})
\Big(
\zeta^{-1} + \left[ -\frac 4{\zeta^2} \hat\Phi(0,0) +
 \frac 2\zeta \left( 1-\frac 2\zeta \right) \Phi(0,0) \right]
\Big).
$$
The $\zeta$ dependence of this asymptotic result is shown in Fig. 4.
The same property is valid for the creation of a finite chain of molecules
with end points $x,x'$, decribed by $H_{xx'}$.

The correlation in $\a$ space $S_{\a\a'}$, describing how
likely it is to change the
type of molecules in the MM, does not depend on the specific pair $\alpha,\a'$
as long as $\a\neq\a'$ because the model connects all $\alpha\neq\a'$
with equal probability:
$$
 S=C(x\alpha ;x\alpha') \approx
 \zeta^{-1}-\zeta^{-2} +
 \frac 1\zeta \Big(1-\frac 6\zeta +\frac 6{\zeta^2}\Big) \hat\Phi(0,0) +
 \frac 2\zeta \Big(1-\frac 1\zeta\Big)^2 \Phi(0,0)
$$
The $N\to\infty$ terms of both correlations $G$ and $S$ agree
but there is a difference in terms of order $1/N$ (s. Fig. 4). It is unlikely
to break an MM and separate the pieces at low and at high density.
This is a consequence of the fact that there are either too few MM
or too many at high densities such that the separation of the pieces
is difficult. It is remarkable that already at densities less than 1
the separation is completely blocked, as indicated by the vanishing $G$
for $n_c\le n\le 1$. Moreover, the substitution of a CM is most likely at
intermediate densities but suppressed at low and high densities.
Again, there is a critical density $n_c'<1$ such that the substitution
is blocked for $n_c'\le n\le 1$. This can be understood as an effect
of interaction: Since there is a exclusion
between molecules of the same type, we can only substitute if there
is no contribution of this molecule from another MM.
At high densities it is very unlikely that a site is not occupied yet
by a CM of a specific type from another MM.

\subsection{Algebraic decay of the correlation function}

To analyse asymptotic behaviour of the correlation function for large distances
$|r-r'|$ of a three-dimensional system, we first have to perform the
$\omega$-integration in Eqs.
(\ref{corr1}) and (\ref{corr2}). This leads to
$$
 \Phi(r-r') = \frac 1N \int  \, \frac{-B(k)(\zeta-1)}{
 \sqrt{ ((\zeta-2)B(k)^2-\zeta)^2 - 4B(k)^2} } \, e^{ik(r-r')}
 \frac{d^{2}k}{(2\pi)^{2}}\; ,
$$
$$
 \hat\Phi(r-r') = \frac 1N \int  \left[ -\frac\zeta{2} +
 \frac{\frac{\zeta^2}2 + B(k)^2 \left( -1+\zeta-\frac{\zeta^2}2 \right)}{
 \sqrt{ ((\zeta-2)B(k)^2-\zeta)^2 - 4B(k)^2} } \right] e^{ik(r-r')}
 \frac{d^{2}k}{(2\pi)^{2}} \; .
$$
The spacially decaying part of the correlation function is given as
$$
 \frac 2{\zeta^2} \left( 1-\frac 1\zeta \right) \Phi(r-r') + \frac 1\zeta
 \left( 1- \frac 2\zeta +\frac 2{\zeta^2} \right) \hat\Phi(r-r') \; .
$$
For small values of $k$ we can assume $B(k) \approx 1$ in the numerators of
$\Phi(r-r')$ and $\hat\Phi(r-r')$. The square root in the
denominators can be approximated by using $B(k)=1-\frac{k^2}2+O(k^4)$:
$$
 \sqrt{ ((\zeta-2)B(k)^2-\zeta)^2 - 4B(k)^2 } = \sqrt{ 4(\zeta-1)k^2+O(k^4) }
 \approx 2 \sqrt{\zeta-1} |k|
$$
Therefore, the asymptotic behavior of the correlation function is
proportional to
$$
 \mbox{const} \, \int \, \frac{e^{ik\cdot (r-r')}}{|k|}
 \frac{d^{2}k}{(2\pi)^{2}} \propto
 \mbox{const} \, \int_0^{2\pi}  \int_0^{\infty} e^{ik|r-r'|\cos\phi}
 dk\ d\phi
$$
$$
 =|r-r'|^{-1} \; \underbrace{\mbox{const} \, \int_0^{2\pi}\int_0^{\infty}  \,
 e^{i\kappa\cos\phi}
 d\kappa\ d\phi}_{=\mbox{const}} \propto |r-r'|^{-1} \; ,
$$
where $k$ has been substituted by $\kappa=k|r-r'|$.
Thus the spatial correlations decay like $1/|r-r'|$ for $d=3$.

\section{Conclusions}

A statistical ensemble of directed MM, constructed with a random sequence
of $N$ different molecules, has been treated for large values of $N$ in terms
of a $1/N$ expansion. We have studied the robustness of the MM with respect
to breaking and replacement of individual CM. Breaking a MM and taking the
pieces apart in space reveals a long-range correlation with an algebraic
decay with the inverse distance. 
The properties in space and in $\a$ are identical 
in the $N\to\infty$ limit but different in $1/N$.

At high densities the broken MM cannot be
separated to arbitrarily large distances due to the interaction between
different MM. The replacement of individual CM is most likely at
intermediate denstities but strongly suppressed at high densities. This
is also a consequence of the repulsive interaction between the molecules
of the same type.


\section*{References}

\begin{thebibliography}{99}
\bibitem{degennes}
de Gennes P G 1979 {\it Scaling Concepts in Polymer Physics}
(Ithaca, NY: Cornell University Press)

\bibitem{li04}
Li Y-Q {\it et al} 2004 {\it Preprint} q-bio.BM/0408024

\bibitem{banavar03}
Banavar J R and Maritan A 2003 {\it Rev. Mod. Phys.} {\bf 75} 23

\bibitem{banavar04}
Banavar J R 2004 {\it Preprint} q-bio.BM/0410031

\bibitem{ziegler}
Ziegler K 1991 {\it Physica A} {\bf 179} 301

\bibitem{berezin}
Berezin F A 1966 {\it Method of Second Quantization} (New York: Academic)
\end{thebibliography}
\end{document}